\documentclass[aps,prl,twocolumn,showpacs,groupedaddress]{revtex4}
\usepackage[dvips]{graphicx}
\usepackage{amssymb}
\usepackage{amsmath}
\usepackage{color}

\begin{document}

\title{Experimental test of the high frequency quantum shot noise theory in a
Quantum Point Contact}
\author{E. Zakka-Bajjani}
\author{J. S\'egala}
\author{F. Portier}
\email{fabien.portier@cea.fr}
\author{P. Roche}
\author{D. C. Glattli}
\altaffiliation[Also at ]{LPA, Ecole Normale Sup\'erieure, Paris.}
\affiliation{Nanoelectronic group, Service de Physique de l'Etat
Condens\'e,\\CEA Saclay, F-91191 Gif-Sur-Yvette, France}
\author{A. Cavanna}
\author{Y. Jin}
\affiliation{CNRS, Laboratoire de Photonique et Nanostructures ,\\
Route de Nozay, F-91460 Marcoussis, France}
\date{\today}
\begin{abstract}

We report on direct measurements of the electronic shot noise of a
Quantum Point Contact (QPC) at frequencies $\nu$ in the range 4-8
GHz. The very small energy scale used ensures energy independent
transmissions of the few transmitted electronic modes and their
accurate knowledge. Both the thermal energy and the QPC
drain-source voltage $V_{ds}$ are comparable to the photon energy
$h\nu$ leading to observation of the shot noise suppression when
$V_{ds}<h\nu /e$. Our measurements provide the first complete test
of the finite frequency shot noise scattering theory without
adjustable parameters.
\end{abstract}

 \maketitle

Pauli's exclusion principle has striking consequences on the
properties of quantum electrical conductors. In an ideal quantum
wire, it is responsible for the quantization of the conductance by
requiring that at most one electron (or two for spin degeneracy)
occupies the regularly time-spaced wave-packets emitted by the
contacts and propagating in the wire \cite{MartinLandauer92}.
Concurrently, at zero temperature, the electron flow is noiseless
\cite{Khlus_87,Lesovik_89} as can be observed in ballistic
conductors \cite{Reznikov95,Kumar96,Delattre06}. In more general
quantum conductors, static impurities diffract the noiseless
electrons emitted by the contacts. This results in a partition of
the electrons between transmitted or reflected states, generating
quantum shot noise \cite{MartinLandauer92,
Khlus_87,Lesovik_89,Buttiker90,Blanter}. However, Pauli's
principle possesses more twists to silence electrons. At finite
frequency $\nu$, detection of current fluctuations in an external
circuit at zero temperature requires emission of photons
corresponding to a finite energy cost $h\nu$ \cite{Lesovik_97}.
 For drain-source contacts biased at voltage $V_{ds}$, a sharp
suppression is expected to occur when the photon energy $h\nu$ is
larger than $eV_{ds}$ as an electron emitted by the source can not
find an empty state in the drain to emit such a photon
\cite{Lesovik_97,Aguado00,Gavish}. Another striking consequence of
Pauli's principle is the prediction of non-classical photon
emission for a conductor transmitting only one or few electronic
modes. It has been shown that in the frequency range $ eV_{ds}/2h
< \nu < eV_{ds}/h$, the population of a photon mode obeys a
sub-Poissonian statistics inherited from the electrons
\cite{SubPoissonianPhotons}. Investigating quantum shot noise in
this high frequency regime using a Quantum Point Contact (QPC) to
transmit few modes is thus highly desirable.

The first step is to check the validity of the above prediction
based on a non-interacting picture of electrons. For 3D or 2D wide
conductors with many quantum channels which are good Fermi
liquids, one expects this non-interacting picture to work well.
Indeed, the $eV_{ds}/h$ singularity has been observed in a 3D
diffusive wire in the shot noise derivative with respect to bias
voltage \cite{Schoelkopf97}. However, for low dimensional systems
like 1D wires or conductors transmitting one or few channels,
electron interactions give non-trivial effects. Long 1D wires
defined in 2D electron gas or Single Wall Carbon Nanotubes become
Luttinger liquids. Long QPCs exhibit a 0.7 conductance anomaly
\cite{Thomas}, and a low frequency shot noise \cite{Roche}
compatible with Kondo physics \cite{Meir}. Consequently, new
characteristic frequencies may appear in shot noise reflecting
electron correlations. Another possible failure of the
non-interacting finite frequency shot noise model could be the
back-action of the external circuit. For high impedance circuits,
current fluctuations implies potential fluctuations at the
contacts \cite{feedback}. Also, the finite time required to
eliminate the sudden drain-source charge build-up after an
electron have passed through the conductor leads to a dynamical
Coulomb blockade for the next electron to tunnel. A peak in the
shot noise spectrum at the electron correlation frequency $I/e$ is
predicted for a tunnel junction connected to a capacitive circuit
\cite{LikharevAverin85}. Other timescales may also be expected
which affect both conductance \cite{BPT93PRL} and noise
\cite{Pedersen98} due to long range Coulomb interaction or
electron transit time. This effects have been recently observed
for the conductance \cite{Gabelli0607} .

The present work aims at giving a clear-cut test of the
non-interacting scattering theory of finite frequency shot noise
using a Quantum Point Contact transmitting only one or two modes
in a weak interaction regime. It provides the missing reference
mark to which further experiments in strong interaction regime can
be compared in the future. We find the expected shot noise
suppression for voltages $\leq h\nu /e$ in the whole 4-8 GHz
frequency range. The data taken for various transmissions
perfectly agree with the finite temperature, non-interacting model
with no adjustable parameter. In addition to provide a stringent
test of the theory, the technique developed is the first step
toward the generation of non-classical photons with QPCs in the
microwave range \cite{SubPoissonianPhotons}. The detection
technique uses cryogenic linear amplification followed by room
temperature detection. The electron temperature much lower than
$h\nu /k_{B}$, the small energy scale used ($eV_{ds} \ll 0.02
E_{F}$) ensuring energy independent transmissions, the high
detection sensitivity, and the absolute calibration allow for
direct comparison with theory without adjustable parameters. Our
technique differs from the recent QPC high frequency shot noise
measurements using on-chip Quantum Dot detection in the 10-150 GHz
frequency range \cite{Onac06}. Although most QPC shot noise
features were qualitatively observed validating this promising
method, the lack of independent determination of the QPC-Quantum
Dot coupling, and the large voltage used from $0.05$ to $0.5
E_{F}$ making QPC transmissions energy dependent, prevent
quantitative comparison with shot noise predictions. However,
Quantum Dot detectors can probe the vacuum fluctuations via the
stimulated noise while the excess noise detected here only probes
the emission noise \cite{Aguado00,Lesovik_97}.

\begin{figure}
\centerline{\includegraphics[angle=-90,width=6cm,keepaspectratio,clip]{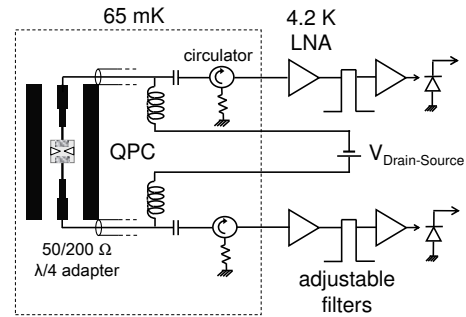}}
\caption{Schematic diagram of the measurement set-up. See text for
details.} \label{Schema.fig}
\end{figure}

The experimental set-up is represented in fig. \ref{Schema.fig}. A
two-terminal conductor made of a QPC realized in a 2DEG in
GaAs/GaAlAs heterojunction is cooled at 65 mK by a dilution
refrigerator and inserted between two transmission lines. The
sample characteristics are a 35 nm deep 2DEG with 36.7
m$^2$V$^{-1}$s$^{-1}$ mobility and 4.4 10$^{15}$ m$^{-2}$ electron
density. Interaction effects have been minimized by using a very
short QPC showing no sign of 0.7 conductance anomaly. In order to
increase the sensitivity, we use the microwave analog of an
optical reflective coating. The contacts are separately connected
to 50 $\Omega$ coaxial transmission lines via two quarter wave
length impedance adapters, raising the effective input impedance
of the detection lines to 200 $\Omega$ over a one octave bandwidth
centered on 6 GHz. The $200  \ \Omega$ electromagnetic impedance
is low enough to prevent dynamical Coulomb blockade but large
enough for good current noise sensitivity. The transmitted signals
are then amplified by two cryogenic Low Noise Amplifiers (LNA)
with $T_{\mathrm{noise}} \simeq 5 \mathrm{K}$. Two rf-circulators,
thermalized at mixing chamber temperature protect the sample from
the current noise of the LNA and ensure a circuit environment
 at base temperature. After
further amplification and eventually narrow bandpass filtering at
room temperature, current fluctuations are detected using two
calibrated quadratic detectors whose output voltage is
proportional to noise power. Up to a calculable gain factor, the
detected noise power contains the weak sample noise on top of a
large additional noise generated by the cryogenic amplifiers. In
order to remove this background, we measure the excess noise
$\Delta S_I(\nu, T, V_{ds}) = S_I(\nu, T, V_{ds})-S_I(\nu, T,
0)$. Practically, this is done by applying a 93 Hz $0$-$V_{ds}$
square-wave bias voltage on the sample through the DC input of a
bias-T, and detecting the first harmonic of the square-wave noise
response of the detectors using lock-in techniques. In terms of
noise temperature referred to the 50 $\Omega$ input impedance, an
excess noise $\Delta S_I(\nu, T,V_{ds})$ gives rise to an
excess noise temperature
\begin{equation}
\Delta T^{50 \Omega}_n(\nu, T, V_{ds}) =
\frac{Z_{\mathrm{eff}}Z_{\mathrm{sample}}^2 \Delta S_I(\nu, T,
V_{ds})}{(2 Z_{\mathrm{eff}}+ Z_{\mathrm{sample}})^2}
.\label{impedancematch.eq}
\end{equation}
Eq. \ref{impedancematch.eq} demonstrates the advantage of
impedance matching : in the high source impedance limit
$Z_{\mathrm{sample}} \gg Z_{\mathrm{eff}}$, the increase in noise
temperature due to shot noise is proportional to
$Z_{\mathrm{eff}}$. Our set up ($Z_{\mathrm{eff}}=200 \Omega$) is
thus four times more efficient than a direct connection of the
sample to standard 50~$\Omega$ transmission lines. Finally, the
QPC differential conductance $G$ is simultaneously measured
through the DC input of the bias-Tee using low frequency lock-in
technique.

\begin{figure}
\centerline{\includegraphics[width=6cm,keepaspectratio,clip,angle=-90]{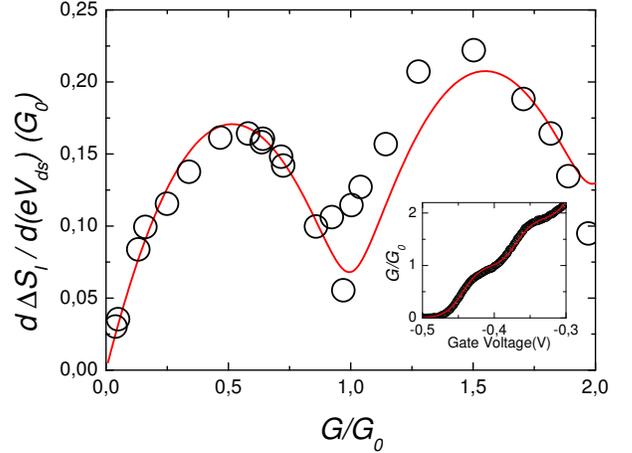}}
\caption{Open circles: $d\Delta S_I / d(eV_{ds})$ deduced from
$\Delta T^{50 \Omega}_n$. Full line : theoretical prediction. The
only fitting parameter is the microwave attenuation. The
experimental uncertainty corresponds to the size of the symbols.
Inset : Open circles : conductance of the QPC as a function of
gate voltage. Solid Line : fit with the saddle point model
\cite{buttiker90}.} \label{T(1-T).fig}
\end{figure}

The very first step in the experiment is to characterize the QPC.
The inset of fig. \ref{T(1-T).fig} shows the differential
conductance versus gate voltage when the first two modes are
transmitted.  As the experiment is performed at zero magnetic
field, the conductance exhibits plateaus quantized in units of
$G_0=2e^2/h$. The short QPC length (80 nm) leads to a conductance
very linear with the low bias voltage used ($\delta G/G \le 6 \%$
for $V_{ds} \le 80 \mu$V for $G \simeq 0.5 \, G_0$). It is also
responsible for a slight smoothing of the plateaus. Each mode
transmission is extracted from the measured conductance (open
circles) by fitting with the saddle point model (solid line)
\cite{buttiker90}.

\begin{figure}[h]
\centerline{\includegraphics[angle=-90,width=7cm,keepaspectratio,clip]{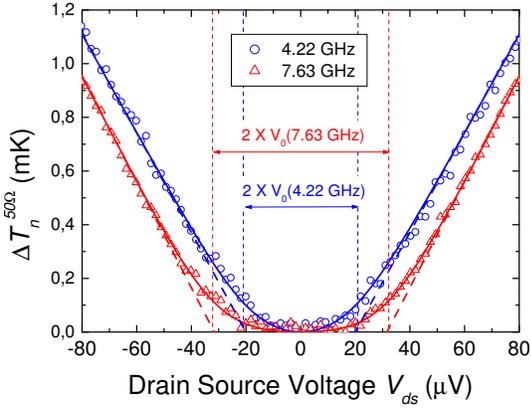}}
\caption{Color Online. Excess noise temperature as a function of
bias voltage, measured at 4.22 GHz (open circles) and 7.63 GHz
(open triangles). The dashed lines
represent the linear fits to the data, %in the [50 $\mu$V, 80 $\mu$V] range for $\vert V_{ds}\vert$,
from which the threshold $V_0$ is deduced. The solid lines
represent the expected excess noise $S_I(\nu,T_e(V_{ds}),
V_{ds})-S_I(\nu, T_e(0), 0)$, using $T_e(V_{ds})$ obtained from
eq. \ref{chauffage.eq}. The frequency dependent coupling is the
only fitting parameter.} \label{deltatn.fig}
\end{figure}

We then set the gate voltage to obtain a single mode at half
transmission corresponding to maximum electron partition ($G
\simeq 0.5 \, G_0$). Fig. \ref{deltatn.fig} shows typical excess
noise measured at frequencies 4.22 GHz and 7.63 GHz and bandwidth
90 MHz and 180 MHz. We note a striking suppression of shot noise
variation at low bias voltage, and that the onset of noise
increases with the measurement frequency. This is in agreement
with the photon suppression of shot noise in a non-interacting
system. The expected excess noise reads
\begin{gather} \Delta
S_I(\nu, T, V_{ds})=  2 G_0 \sum_i D_i(1-D_i)
 \left( \displaystyle{\frac{h \nu -eV_{ds}}{e^{(h \nu -eV_{ds})/k_{\mathrm{B}}T}-1}}  \right. \nonumber \\
\left.  + \displaystyle{\frac{h \nu +eV_{ds}}{e^{(h \nu
+eV)/k_{\mathrm{B}}T }-1}} - \displaystyle{\frac{2 h \nu}{e^{h \nu
/k_{\mathrm{B}}T }-1}} \right).
 \label{DeltaTn.Eq}
\end{gather}

It shows a zero temperature singularity at $eV_{ds}=h\nu$ :
$\Delta S_I(\nu, T, V_{ds}) = 2 G_0 \sum_i D_i(1-D_i)(eV_{ds}-h
\nu)$  if $eV_{ds}
> h \nu$ and 0 otherwise. At finite temperature, the singularity
is thermally rounded. At high bias ($eV_{ds} \gg h \nu$,
$k_{\mathrm{B}} T$), equation \ref{DeltaTn.Eq} gives an excess
noise
\begin{eqnarray}
& \Delta S_I(\nu, T, V_{ds}) =2 G_0 \sum_i D_i(1-D_i) \left( eV_{ds}
- eV_0 \right) & \label{shot-asymptote.eq}\\
& \mbox{with} \qquad eV_0=h \nu \coth \left(h \nu / 2
k_{\mathrm{B}}T \right).& \label{V0.eq}
\end{eqnarray}

\begin{figure}[h]
\centerline{\includegraphics[angle=-90,width=7cm,keepaspectratio,clip]{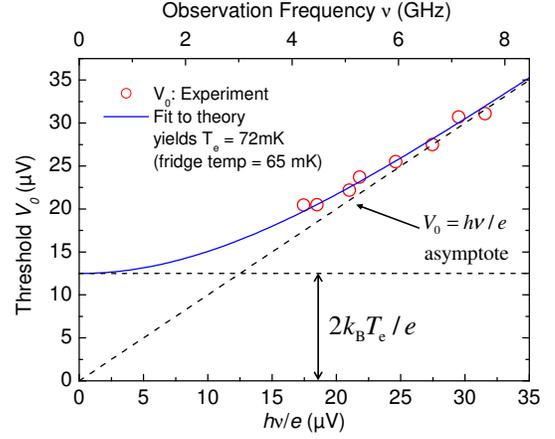}}
\caption{Onset $V_0$  as a function of the observation frequency.
The experimental uncertainty corresponds to the size of the
symbols. The dashed lines correspond to the low ($e V_0=2
k_{\mathrm{B}} T$) and high ($e V_0= h \nu$) frequency limits, and
the solid line is a fit to theory, with the electronic temperature
as only fitting parameter.} \label{V0.fig}
\end{figure}

In the low frequency limit, the threshold $V_0$ characterizes the
transition between thermal noise and shot noise ($eV_0=2
k_{\mathrm{B}}T$), whereas in the low temperature limit, it marks
the onset of photon suppressed shot noise ($eV_0=h\nu$). As shown
on fig. \ref{deltatn.fig}, $V_0$ is determined by the intersection
of the high bias linear regression of the measured excess noise
and the zero excess noise axis. Fig. \ref{V0.fig} shows $V_0$ for
eight frequencies spanning in the 4-8 GHz range for $G \simeq 0.5
\, G_0$ . Eq. \ref{V0.eq} gives a very good fit to the
experimental data. The only fitting parameter is the electronic
temperature $T_{\mathrm{e}}$ = 72 mK, very close to the fridge
temperature $T_{\mathrm{fridge}}$ = 65 mK.  We will show that
electron heating can account for this small discrepancy.

To get a full comparison with theory, we now investigate the
influence of the transmissions of the first two electronic modes
of the QPC. To do so, we repeat the same experiment at fixed
frequency (here we used a 5.4-5.9 GHz filter) for different sample
conductances. The noise suppression at $V_{ds} \le h\nu / e$ is
the only singularity we observe, independently of the QPC
conductance $G$. Fig. \ref{T(1-T).fig} shows the derivative with
respect to $eV_{ds}$ of the excess noise $d\Delta S_I /
d(eV_{ds})$ deduced from the excess noise temperature measured
between 50 $\mu$V and 80 $\mu$V. This energy range is chosen so
that $eV_{ds}$ is greater than $h \nu$ by at least $5
k_{\mathrm{B}}T_{\mathrm{fridge}}$ over the entire frequency
range. The data agree qualitatively with the expected $D(1-D)$
dependence of pure shot noise, showing maxima at conductances
$G=0.5 \, G_0$, and $G=1.5 \, G_0$, and minima at conductances
$G=G_0$ and $G=2\, G_0$. The short QPC is responsible for the non
zero minima as, when the second mode starts to transmit electrons,
the first one has not reached unit
 transmission (inset
 of fig. \ref{T(1-T).fig}). However, eq. \ref{DeltaTn.Eq}
 is not compatible with a
 second maximum higher than the first one, which is due to electron heating. The
dimensions of the 2-DEG being much larger than the
electron-electron energy relaxation length, but much smaller than
electron-phonon energy relaxation length, there is a gradient of
electronic temperature from the QPC to the ohmic metallic contacts
assumed at the fridge temperature. Combining the dissipated power
$IV_{ds}$ with the Wiedemann-Franz law, one gets
\cite{Kumar96,Steinbach96}
\begin{equation}
T_{\mathrm{e}}^2=T_{\mathrm{fridge}}^2+\frac{24}{\pi^2}\frac{G}{G_m}\left(1+\frac{2G}{G_m}\right)
\left(\frac{eV_{ds}}{2 k_{\mathrm{B}}}\right)^2 \label{chauffage.eq}
\end{equation}
where $G_m$ stands for the \emph{total} conductance of the 2D
leads, estimated from measurements to be 12 mS $\pm 20 \%$. The
increased noise temperature is then due to both shot noise and to
the increased thermal noise. For a fridge temperature of 65 mK and
$G=G_0/2$, the electronic temperature will increase from 69 mK to
77 mK as $V_{ds}$ increases from 50 $\mu$V to 80 $\mu$V. This
accounts for the small discrepancy between the fridge temperature
and the electron temperature deduced from the variation of $V_0$
with frequency. As $G$ increases, the effect is more important, as
can be seen both in fig. \ref{T(1-T).fig} and eq.
\ref{chauffage.eq}. The solid line in figure \ref{T(1-T).fig}
gives the average derivative with respect to $eV_{ds}$ of the
total expected excess noise $S_I(\nu, T_e(V_{ds}),
V_{ds})-S_I(\nu,T_e(0), 0)$, using the attenuation of the signal
as a free parameter. The agreement is quite satisfactory, given
the accuracy of the saddle point model description of the QPC
transmission. We find a 4.7 dB attenuation, which is in good
agreement with the expected 4 $\pm 1$ dB deduced from calibration
of the various elements of the detection chain. Moreover, the
voltage dependent electron temperature obtained from eq.
\ref{chauffage.eq} can also be used to evaluate $S_I(\nu,
T_e(V_{ds}), V_{ds})-S_I(\nu,T_e(0), 0)$ as a function of $V_{ds}$
at fixed sample conductance $G = 0.5 \, G_0$. The result, as shown
by the solid lines of fig. \ref{deltatn.fig}, is in excellent
agreement with experimental observations.

In conclusion, we performed the first direct measurement of the
finite frequency shot noise of the simplest mesoscopic system, a
QPC. Accurate comparison of the data with non-interacting shot
noise predictions have been done showing perfect quantitative
agreement. Even when a single mode is transmitted, no sign of
deviation related to interaction was found, as expected for the
experimental parameters chosen for this work. We have also shown
that accurate and reliable high frequency shot noise measurements
are now possible for conductors with impedance comparable to the
conductance quantum. This opens the way to high frequency shot
noise characterization of Carbon Nanotubes, Quantum Dots or
Quantum Hall samples in a regime where microscopic frequencies are
important and will encourage further theoretical work in this
direction. Our set-up will also allow to probe the statistics of
photons emitted by a phase coherent single mode conductor.
\begin{acknowledgments}
It is a pleasure to thank D. Darson, C. Ulysse, P. Jacques and C.
Chaleil for valuable help in the construction of the experiments,
P. Roulleau for technical help, and X. Waintal for useful
discussions.
\end{acknowledgments}


\begin{references}

\bibitem{MartinLandauer92} T. Martin and R. Landauer, Phys.
Rev. B \textbf{45}, 1742 (1992)

\bibitem{Khlus_87}  V. A. Khlus, Zh. Eksp. Teor. Fiz. \textbf{93} (1987) 2179 [Sov.
Phys. JETP \textbf{66} (1987) 1243].

\bibitem{Lesovik_89} G. B. Lesovik, Pis'ma Zh.
Eksp. Teor. Fiz. \textbf{49} (1989) 513 [JETP Lett. \textbf{49}
(1989) 592].

\bibitem{Reznikov95} M. Reznikov, \emph{et
al.},  Phys. Rev. Lett. {\bf 75}, 3340 (1995);

\bibitem{Kumar96} A. Kumar \emph{et al.}, Phys. Rev. Lett. {\bf 76}, 2778 (1996).

\bibitem{Delattre06} L. Hermann \emph{et al.},
arXiv:cond-mat/0703123v1.

\bibitem{Buttiker90} M. B\"{u}ttiker, Phys. Rev. Lett. \textbf{65}, 2901 (1990)

\bibitem{Blanter} Y. M. Blanter and M.
B\"{u}ttiker, Phys. Rep. \textbf{336}, 1 (2000).

\bibitem{Lesovik_97} G.B. Lesovik, R. Loosen, JETP Lett. \textbf{65}, 295
(1997). Here is made the distinction between emission noise
$S_{I}(\nu)=\int_{-\infty}^{+\infty}\langle I(0)I(\tau )\rangle
e^{i2\pi \nu \tau} d \tau$ and stimulated noise $S_{I}(-\nu)$.
While observation of the later requires excitation of the sample
by external sources, for a zero temperature external circuit, only
$S_{I}(\nu)$ should be observed. For an earlier high frequency
shot noise derivation not making the distinction between
$S_{I}(\nu)$ and $S_{I}(-\nu)$, see
Ref.\cite{Khlus_87,Lesovik_89}.

\bibitem{Aguado00} R. Aguado and L. P. Kouwenhoven, Phys. Rev. Lett.
\textbf{84}, 1986 (2000);

\bibitem{Gavish} U. Gavish, Y. Levinson, Y. Imry,
Phys. Rev. B \textbf{62}, R10637 (2000); M. Creux, A. Crepieux, Th.
Martin, Phys. Rev. B \textbf{74} 115323 (2006).

\bibitem{SubPoissonianPhotons}  C. W. J Beenakker and H. Schomerus, Phys. Rev. Lett. \textbf{86}, 700
(2001); J. Gabelli, \emph{et al.},  Phys. Rev. Lett. \textbf{93},
056801 (2004); C. W. J. Beenakker and H. Schomerus Phys. Rev.
Lett. \textbf{93}, 096801 (2004).

\bibitem{Schoelkopf97} R. J. Schoelkopf \emph{et al.}, Phys. Rev. Lett. \textbf{78}, 3370 (1997).

\bibitem{Thomas} K. J. Thomas \emph{et al.}, Phys. Rev. Lett. \textbf{77}, 135 (1996); K. J.
Thomas \emph{et al.}, Phys. Rev. B \textbf{58}, 4846 (1998).

\bibitem{Roche} P. Roche \emph{et al.}, Phys. Rev. Lett. \textbf{93}, 116602 (2004); L. DiCarlo \emph{et al.},
Phys. Rev. Lett. \textbf{97}, 036810 (2006).

\bibitem{Meir} A. Golub, T. Aono, and Y. Meir Phys. Rev. Lett. \textbf{97}, 186801 (2006)

\bibitem{feedback}     B. Reulet, J. Senzier, and D. E. Prober, Phys. Rev. Lett. \textbf{91},
196601 (2003); M. Kindermann, Yu. V. Nazarov, and C. W. J.
Beenakker Phys. Rev. B \textbf{69}, 035336 (2004).

\bibitem{LikharevAverin85} D.V. Averin and K.K. Likharev, J. Low
Temp.Phys. \textbf{62} 345 (1986).


\bibitem{BPT93PRL}
M. B\"{u}ttiker, H. Thomas, and A. Pr\^{e}tre, Phys. Lett. {\bf
A180}, 364 (1993); M. B\"uttiker, A. Pr\^etre, H. Thomas,
\emph{Phys. Rev. Lett.} \textbf{70}, 4114 (1993)


\bibitem{Pedersen98} M. H. Pedersen, S. A. van Langen, and M. Buttiker,
Phys. Rev. B \textbf{57} (1998) 1838.


\bibitem{Gabelli0607} J. Gabelli \emph{et al.}, \emph{Science} \textbf{313}, 499 (2006).
J. Gabelli \emph{et al.}, Phys. Rev. Lett. \textbf{98}, 166806
(2007)


\bibitem{Onac06} E. Onac \emph{et al.}
Phys. Rev. Lett. \textbf{96}, 176601 (2006). The experimental
onset in $V_{ds}$ for the emission of high frequency shot noise
was larger than expected  ($V_{ds} \simeq 5\times h\nu /e$). After
submission of this work, Gustavson {\textit et al.} reported on a
double quantum dot on-chip detector, yielding to a more
quantitative agreement with theory (arXiv:0705.3166v1).


\bibitem{buttiker90} M. B{\"{u}}ttiker Phys. Rev. B \textbf{41}, 7906-7909
(1990).

\bibitem{Steinbach96} A. H. Steinbach, J. M. Martinis, and M. H. Devoret Phys. Rev. Lett. \textbf{76}, 3806 (1996)

\end{references}
\end{document}